\documentclass[aps,pre,twocolumn,showpacs]{revtex4}
\usepackage{epsfig}\usepackage{times}
\begin{document}

\title{Cyclical interactions with alliance-specific heterogeneous invasion rates}

\author{Matja{\v z} Perc$^\star$, Attila Szolnoki$^\dagger$, and Gy\"orgy Szab\'o$^\dagger$}\affiliation{$^\star$Department of Physics, Faculty of Education, University of Maribor, Koro{\v s}ka cesta 160, SI-2000 Maribor, Slovenia\\$^\dagger$Research Institute for Technical Physics and Materials Science, P.O. Box 49, H-1525 Budapest, Hungary}

\begin{abstract}
We study a six-species Lotka-Volterra type system on different two-dimensional lattices when 
each species has two superior and two inferior partners. The invasion rates from predator sites 
to a randomly chosen neighboring prey's site depend on the predator-prey pair, whereby cyclic 
symmetries within the two three-species defensive alliances are conserved. Monte Carlo 
simulations reveal an unexpected non-monotonous dependence of alliance survival on the 
difference of alliance-specific invasion rates. This behavior is qualitatively reproduced 
by a four-point mean-field approximation. The study addresses fundamental problems of 
stability for the competition of two defensive alliances and thus has important implications 
in natural and social sciences.
\end{abstract}


\maketitle

Cyclical interactions are simple yet fascinating and powerful examples of 
evolutionary processes \cite{hofbauer_98}, able to provide insights into the 
intriguing mechanisms of Darwinian selection \cite{maynard_n73} as well as structural 
complexity \cite{watt_je47} and pre-biotic evolution \cite{rasmussen_s04}. The simplest 
non-trivial food web describing such cyclical interactions is formed by three species that 
have relationships analogous to the well-known rock-scissors-paper (RSP) game, where 
strategies form a closed loop of dominance. 
Real-life examples of such interactions include the mating strategy 
of side-blotched lizards \cite{sinervo_n96}, overgrowths by marine sessile 
organisms \cite{burrows_mep98}, and competition among different strains of 
bacteriocin-producing bacteria \cite{szabo_pre01a}. Cycles are also common 
in the context of evolutionary game theory \cite{nowak06}, where strategic 
complexity \cite{hauert_s02,traulsen_pre04} often leads to RSP type of 
dominance between different strategies \cite{semmann_n03}.

Several theoretical aspects of multi-species cyclical dominance have already been studied in 
detail. For example, it has been established that three species in a cyclic dominance 
exhibit self-organizing behavior on the spatial grid \cite{tainaka_prl89,frachebourg_pre96},
whereby similar observations can be made also for system that incorporate more than three 
species, provided their total number does not exceed fourteen \cite{frachebourg_jpa98}. Phase 
transitions and selection have also been studied in the predator-prey models allowing
motion throughout inhabitable vacant sites \cite{szabo_pre04a,he_ijmpc05}. Especially reticulate 
six-species models with mutation \cite{szabo_pre01b} and local mixing \cite{szabo_jpa05} have 
recently been studied rigorously, reporting the spontaneous emergence of defensive alliances and 
numerous stable spatial distributions as well as pertaining phase transitions in dependence on 
the topology of underlying food webs. In these systems the number of possible stationary 
states increases rapidly with the number of species because the solutions of subsystems 
(some species are missing) are also solutions of the whole system. The final 
stationary state can be determined by the competition of subsystem solutions (defensive alliances) 
that are characterized by their composition and spatio-temporal distribution of 
species. Consequently, the understanding of systems with many species requires the systematic analysis
of all their subsystems. In the present work we will
study a predator-prey model that can be considered as a six-species subsystem of strains
of bacteria using two types of toxins and anti-toxins in their warfare \cite{szabo_pre01a}.

Besides increasing the number of species, the complexity of models can be enhanced also by the 
introduction of heterogeneous invasion rates between interacting individuals. Differences in 
invasion rates might affect the proportions of participating species in the 
habitat \cite{frean_prsb01} 
as well as geometrical features of patterns on the spatial grid \cite{szabo_pre02a}. Moreover, 
it has recently been discovered that multi-species models of cyclical interactions comprising 
an even number of individuals are very sensitive to the independent variation of invasion 
rates \cite{sato_amc02}. The introduction of heterogeneous invasion rates also 
raises questions about the relevance of defensive alliances. 

\begin{figure}
\centerline{\epsfig{file=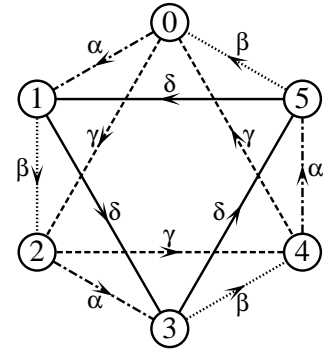,width=4.5cm}}\caption{\label{fig:fw}Food web of the 
studied predator-prey model. Arrows point from predators towards prey with heterogeneous 
invasion rates specified along the edges.}
\end{figure}

Presently, we thus study a reticulate six-species predator-prey model where each site $i$ of 
the square lattice is occupied by an individual belonging to one of the six species. Their 
corresponding distribution is given by a set of site variables $s_i = 0,\dots, 5$. The 
predator-prey relations and the corresponding invasion 
rates ($0 < \alpha, \beta, \gamma, \delta <1$) are defined by the food web presented 
in Fig.~\ref{fig:fw}. For this choice of parameterization the two subsystems consisting of 
odd and even labeled species are equivalent to the thoroughly studied rock-scissors-paper 
game and the system remains unchanged under cyclic permutation ($s \to s + 2$ modulo 6) 
of species.

In case of homogeneous rates ($\alpha=\beta=\gamma=\delta=1$) the system has two 
equivalent three-species states [denoted by $(0+2+4)$ and $(1+3+5)$] exhibiting a 
self-organizing pattern maintained by cyclic invasions. These state are called defensive 
alliances because their members protect each other cyclically against the external 
invaders \cite{szabo_pre01b,szabo_jpa05}. Consider for example the case when the 
species $0$ invades allies $(1+3+5)$, in particular by attacking species $1$. Its 
intention is immediately disabled by the species $5$ that is superior to both $0$ and $1$. Thus, 
the intruder $0$ is quickly abolished from the $(1+3+5)$ domain by the very same species $5$ that 
dominates species $1$ within the alliance. The same reasoning applies for all 
other possible attempts of non-allied species to invade a defensive alliance. Importantly, in 
this mechanism the proper spatio-temporal distribution of species plays a crucial 
role (the mean-field approximation cannot reproduce this feature). 

By introducing 
alliance-specific heterogeneities in the invasion rates, we are capable of analyzing the 
competition between the defensive alliances. For example, we can study what happens when 
one of the associations is more aggressive towards the other ($\alpha \neq \beta$) or when 
the internal mechanism fails to assure flawless protection against the 
invaders ($\gamma  \neq \delta$). In order to address these two issues systematically, 
it appears reasonable to introduce two parameters that, due to symmetries in the food web, 
uniquely determine the stationary state of the system. Particularly, 
let $G = \beta - \alpha$ and $H = \gamma - \delta$ where  $H, G \in [-1, 1]$. Note that 
the system behavior becomes trivial in two quadrants of the $H-G$ parameter space because 
the favored defensive alliance is supported by both mentioned mechanisms, thus leading to 
its undisputed dominance on the spatial grid. If $G < 0$ ($\beta < \alpha$) 
and $H > 0$ ($\gamma > \delta$) allies $(0+2+4)$ receive a two-fold advantage. 
First, due to $\beta < \alpha$ $(0+2+4)$ invade non-members faster than individuals 
in the competing alliance. Second, due to $\gamma > \delta$ the internal invasions within 
$(0+2+4)$ are faster than those in $(1+3+5)$, whereby faster internal cyclic 
invasions uphold a more effective protection shield against the external invaders by assuring 
a prompt response to a potential attack. An interesting competition emerges only 
if $G > 0$ and $H > 0$ (or equivalently if $G < 0$ and $H < 0$), which we are going to 
explore next. Due to the symmetry of the problem our analysis will be constrained to the 
parameter space spanning over $H, G \in [0, 1]$. 

We perform Monte Carlo (MC) simulations of the introduced six-species cyclical interaction 
model on the $L \times L$ spatial grid. 
Initially the six species are randomly distributed. The elementary steps are the following.
First, two nearest neighbors are chosen at random, and second, 
if the two neighboring species form a predator-prey pair (species connected by an arrow in Fig.~\ref{fig:fw}) 
the prey is killed with the rate specified along the arrow and an offspring of the predator occupies the prey's site. 
On the other hand, if the two randomly chosen species form a neutral pair (species not connected by an arrow 
in Fig.~\ref{fig:fw}), 
or if both are identical, the second step dictates no action (nothing happens) and the MC simulation proceeds 
with executing step one.
In accordance with the random sequential update, each individual is selected once on average during a particular 
Monte Carlo step (MCS). In order to characterize the stationary state we define the order 
parameter $m = \rho_1 + \rho_3 + \rho_5 - \rho_0 - \rho_2 - \rho_4$, 
whereby  $\rho_s$ ($s=0, \ldots, 5$) denotes the fraction of species $s$ on the spatial grid. 
Here $m = 1$ corresponds to the complete dominance of the alliance $(1+3+5)$. On the other 
hand, $m = -1$ indicates the absolute authority of the association $(0+2+4)$.

\begin{figure}
\centerline{\epsfig{file=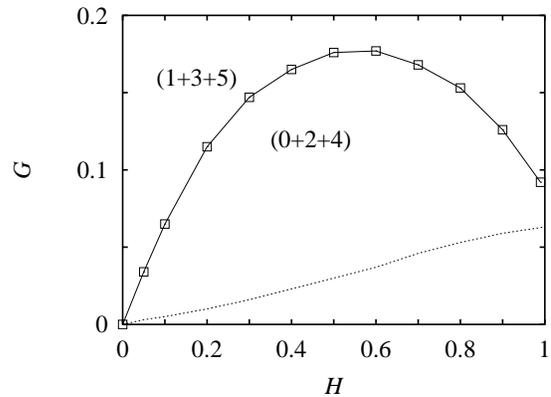,width=8cm}}
\caption{\label{fig:nm}Phase separation line dividing the two pure phases, characterized either by 
the exclusive dominance of the alliance $(0+2+4)$ or $(1+3+5)$, in dependence on $H$ and $G$ ($\beta = \gamma = 1$). 
Linked squares were obtained by MC simulations while the dashed line results from the four-point 
cluster mean-field approximation.}
\end{figure}

We start by setting $G$ and $H$ equal to zero, whereby both alliances have an equal chance of 
eventually dominating the spatial grid via a domain growing process. In accordance with the 
above discussion of the studied model, it is reasonable to expect that as soon as $G$ rises above 
zero (keeping $H=0$) allies $(1+3+5)$ are favored as their members invade non-allied species 
more successfully ($\alpha < \beta$). Thus, $m = 1$ in the stationary state, meaning that the 
system evolves into the three-species self-organizing phase $(1+3+5)$ as soon as members 
of $(0+2+4)$ die out. However, the advantage of $(1+3+5)$ given by $G > 0$ 
can be compensated by choosing sufficiently large values of $H$, as shown in Fig.~\ref{fig:nm}. 
As argued above, if $H$ rises above zero the internal invasions within allies $(1+3+5)$ 
slow down in comparison to $(0+2+4)$, thus decreasing the effectiveness 
of the protection of the odd alliance and in turn nullifying its advantage given 
by $G > 0$. 

Strikingly though, results of MC simulations shown in Fig.~\ref{fig:nm} reveal a 
non-monotonous phase diagram in dependence on $H$. In particular, the advantage of 
allies $(1+3+5)$ again increases if $H$ approaches 1 [the internal invasion rate ($\delta$) 
vanishes]. As $\delta \to 0$ allies $(1+3+5)$ essentially stop to invade each 
other within the alliance.
In this case ($\delta = 0$) an external species (e.g. 2) can survive in the bulk
of its neutral pair (e.g. in the domain of species 5).
As a result, the order parameter $m$ remains below 1
forming a frozen state after achieving dominance. The mechanism 
behind this interesting phenomenon is subtle and cannot be 
grasped at a glance. The above analysis was made also for some other fixed values 
of $\beta$ and $\gamma$ yielding essentially identical results.

Evidently, the mean-field analysis can be applied as an analytical tool to study the behavior 
of the proposed predator-prey model. Unfortunately, the resulting master equations fail to 
confirm above results. Namely, the solution for the fractions of species predicts a total 
dominance of allies $(1+3+5)$ if $G > 0$ irrespective of $H$. This shortage of the classical 
mean-field approximation may be eliminated by applying the extended versions of dynamical 
mean-field theory that proved to be very appropriate for obtaining qualitatively correct phase 
diagrams for several non-equilibrium systems. The improved approach involves finding a 
hierarchy of evolution equations for the configurational probabilities on $k$-site clusters, 
where $k$ characterizes the level of approximation (for details, 
see e.g. \cite{dickman_pla87,szabo_cm06}). Nonetheless, the application of the 
method at the two-point level is still unable to account for the incursion of 
allies $(0+2+4)$ into the $G > 0$ region if $H > 0$. The first level that supports our 
above conjectures is the four-point level, of which the solution is displayed in 
Fig.~\ref{fig:nm}. The fact that results of MC simulations differ from the four-point 
cluster mean-field approximation suggests that the unexpected non-monotonous dependence 
is heavily routed in the short-range correlation of spatial distribution, which cannot be 
captured adequately by the ansatz of a four-point level approximation. 

To obtain a better understanding of the spatial dynamics behind the phase diagram 
in Fig.~\ref{fig:nm} we examine the nature of phase transitions between $m = -1$ and $m = 1$. 
Although the two competing effects of parameters $G$ and $H$ suggest that indeed a fine-tuning 
towards a stable co-existence of both alliances may be possible, thus assuring a rich diversity 
of species on the spatial grid, the reality is quite different. In fact, results 
in Fig.~\ref{fig:nm} suggest that phase transitions between the two pure phases are extremely 
sharp, and thus a stable co-existence of the two defensive alliances on the spatial grid is 
not feasible. Although we have only numerical arguments the extensive 
calculations performed in order to come to this conclusion leave extremely little room for 
alternatives.

\begin{figure}
\centerline{\epsfig{file=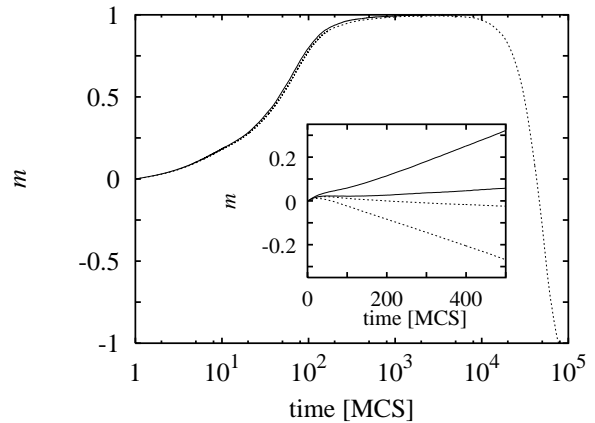,width=8.5cm}}
\caption{\label{fig:sep}Temporal evolution of the order parameter for $H=0.2$ and different 
values of $G$. Solid line depicts the temporal evolution obtained for $G=0.116$, while the 
dashed line corresponds to $G=0.114$. Inset shows the evolution of the order parameter from 
prepared initial states (see text for details). Lines in the inset correspond 
to $G = 0.130, 0.117, 0.113$ and $0.110$ from top to bottom.}
\end{figure}

Finally, it is instructive to examine the temporal evolution of the order parameter in the 
close vicinity of the phase boundary. Figure \ref{fig:sep} illustrates that, at the beginning, 
the fraction of even labeled species decreases on both sides of the phase separation line. 
However, while above the phase boundary members of the even alliance go extinct, 
below the phase boundary allies $(0+2+4)$ are able to fully recover although only a minute portion 
(as little as $0.005$) of the spatial grid is occupied by its members, eventually yielding $m = -1$. 
We argue that this remarkable and unusual behavior, which is intimately 
linked also with the non-monotonous phase diagram presented in Fig.~\ref{fig:nm}, is related 
to the formation of a suitable boundary layer that modifies the interaction between both 
defensive alliances and ultimately tosses the dominance in favor of either $(1+3+5)$ or $(0+2+4)$. 
It is remarkable that a similar temporal evolution was reported in a group selection 
model \cite{silva_epjb99} 
where the altruistic species almost reach extinction before taking over the population. 
However, as described above, in our case the spatiality is a fundamental ingredient since the recovery 
of the even alliance, occurring due to its dynamical benefit, is an interface driven domain growing process.

To confirm the importance of the boundary layer between both defensive alliances, we 
perform a stability analysis of the interface via MC simulations of the system with specially 
prepared initial conditions. In particular, we start the simulations with sharp boundaries that 
separate regions of initially randomly distributed even and odd labeled species. Within 
a $3200 \times 3200$ spatial grid we were able to set up $64$ such straight boundary 
layers (without them interfering with each other) and monitored in which direction the 
interface moved in dependence on $G$ placing the system below or above the phase boundary 
shown in Fig.~\ref{fig:nm}. Additionally, results were averaged over $10$ consecutive runs 
to minimize unwanted fluctuations. The inset in Fig.~\ref{fig:sep} shows the results for two 
values of $G$ below and two above the phase separation line. Note that $m$ measures the 
difference of areas on both sides of the initially sharp boundary between even and odd labeled 
species, thus uniquely determining also its spreading direction. Evidently, if $G$ is set above 
the phase separation line in Fig.~\ref{fig:nm} the boundary layer moves towards the area 
of $(0+2+4)$, thus foretelling an imminent dominance of $(1+3+5)$. Conversely, if $G$ is set 
below the phase separation line the boundary layer moves towards $(1+3+5)$, marking the advent 
of dominance of $(0+2+4)$ although the total fraction of its members on the spatial 
grid at that time might be minute. Hence, if the evolution of the system is initialized from a 
completely random initial distribution of species on the spatial grid, members of $(1+3+5)$ 
can utilize their advantage given by $G > 0$ and rapidly start their conquest. However, 
as time goes by the seemingly defeated $(0+2+4)$ may self-organize into small clusters on 
the spatial grid, and if $G$ and $H$ are set appropriately, start to win back lost ground via 
growth along the interfaces, as explained above and presented in Fig.~\ref{fig:sep}. 

\begin{figure}
\centerline{\epsfig{file=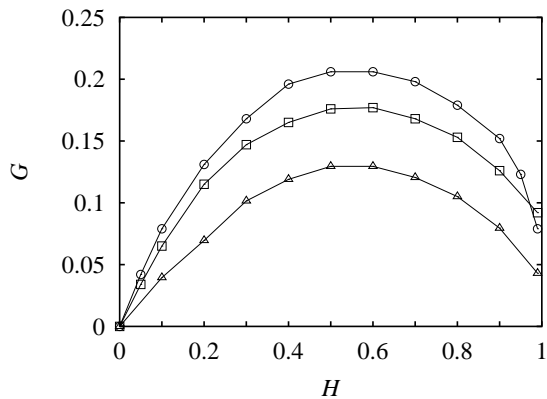,width=8cm}}
\caption{\label{fig:comp}Phase boundaries between the stationary states $(0+2+4)$ and $(1+3+5)$ 
for different lattice structures. Linked symbols depict phase separation lines for the honeycomb 
(circles), square (squares), and triangle (triangles) lattice.}
\end{figure}

In order to check the robustness of the above-described behavior the present predator-prey 
model was also studied on the honeycomb and triangular lattices having different coordination 
numbers. Figure \ref{fig:comp} clearly shows that the honeycomb lattice additionally pronounces 
the non-monotonous variation of the phase boundary. Conversely, the incursion of allies $(0+2+4)$ 
is less emphasized by the triangular lattice, which may be related to the increased coordination 
number, bringing the MC simulations closer to the mean-field behavior. Results presented in 
Fig.~\ref{fig:comp} further stress the importance of spatiality and with it related distribution 
of species and resulting boundary layers among defensive alliances on the spatial grid.    

In sum, we have studied a six-species predator prey model with special heterogeneous invasion 
rates that were introduced in accordance with the two spontaneously emerging defensive alliances. 
We have shown that an increased aggression towards non-allied species could be tamed by decreasing 
the willingness of members to attack individuals within the alliance itself. Remarkably though, if 
individuals completely sized to invade its own allied members the benefit of increased aggression 
towards non-allied species again increased. These two facts resulted in a non-monotonous 
dependence of alliance survival on the difference of alliance-specific invasion rates, which we 
attributed to the underlying spatial dynamics of the system. We also discovered that despite the 
ability of fine-tuning two system parameters, a stable state enabling the co-existence of both 
defensive alliances on the spatial grid is not possible, thus resulting in sharp phase transitions 
between the two absorbing states. However, the newly introduced alliance-specific heterogeneous 
invasion rates might yet prove valuable by discovering new ways of assuring biodiversity, either 
within generalized Lotka-Volterra models \cite{malcai_pre02}, via the impact of 
stochasticity \cite{reichenbach_pre06}, or oscillatory mechanisms \cite{kowalik_pre02}. 
We hope that the study will prove vital for the understanding of the robustness 
of alliance formation, their competition, and for the effect of spatial structures on the evolution of food webs 
in real systems, which appear to range from strains of bacteria to man made economic systems.

This work was supported by the Hungarian National Research 
Fund (T-47003), the European Science Foundation (COST P10), 
and the Slovenian Research Agency (Z1-9629).


\begin{thebibliography}{27}
\expandafter\ifx\csname natexlab\endcsname\relax\def\natexlab#1{#1}\fi
\expandafter\ifx\csname bibnamefont\endcsname\relax
  \def\bibnamefont#1{#1}\fi
\expandafter\ifx\csname bibfnamefont\endcsname\relax
  \def\bibfnamefont#1{#1}\fi
\expandafter\ifx\csname citenamefont\endcsname\relax
  \def\citenamefont#1{#1}\fi
\expandafter\ifx\csname url\endcsname\relax
  \def\url#1{\texttt{#1}}\fi
\expandafter\ifx\csname urlprefix\endcsname\relax\def\urlprefix{URL }\fi
\providecommand{\bibinfo}[2]{#2}
\providecommand{\eprint}[2][]{\url{#2}}

\bibitem[{\citenamefont{Hofbauer and Sigmund}(1998)}]{hofbauer_98}
\bibinfo{author}{\bibfnamefont{J.}~\bibnamefont{Hofbauer}} \bibnamefont{and}
  \bibinfo{author}{\bibfnamefont{K.}~\bibnamefont{Sigmund}},
  \emph{\bibinfo{title}{Evolutionary Games and Population Dynamics}}
  (\bibinfo{publisher}{Cambridge University Press},
  \bibinfo{address}{Cambridge}, \bibinfo{year}{1998}).

\bibitem[{\citenamefont{Maynard~Smith and Price}(1973)}]{maynard_n73}
\bibinfo{author}{\bibfnamefont{J.}~\bibnamefont{Maynard~Smith}}
  \bibnamefont{and} \bibinfo{author}{\bibfnamefont{G.~R.} \bibnamefont{Price}},
  \bibinfo{journal}{Nature} \textbf{\bibinfo{volume}{246}}, \bibinfo{pages}{15}
  (\bibinfo{year}{1973}).

\bibitem[{\citenamefont{Watt}(1947)}]{watt_je47}
\bibinfo{author}{\bibfnamefont{A.~S.} \bibnamefont{Watt}}, \bibinfo{journal}{J.
  Ecol.} \textbf{\bibinfo{volume}{35}}, \bibinfo{pages}{1}
  (\bibinfo{year}{1947}).

\bibitem[{\citenamefont{Rasmussen et~al.}(2004)\citenamefont{Rasmussen, Chen,
  Deamer, Krakauer, Packard, Stadler, and Bedau}}]{rasmussen_s04}
\bibinfo{author}{\bibfnamefont{S.}~\bibnamefont{Rasmussen}},
  \bibinfo{author}{\bibfnamefont{L.}~\bibnamefont{Chen}},
  \bibinfo{author}{\bibfnamefont{D.}~\bibnamefont{Deamer}},
  \bibinfo{author}{\bibfnamefont{D.~C.} \bibnamefont{Krakauer}},
  \bibinfo{author}{\bibfnamefont{N.~H.} \bibnamefont{Packard}},
  \bibinfo{author}{\bibfnamefont{P.~F.} \bibnamefont{Stadler}},
  \bibnamefont{and} \bibinfo{author}{\bibfnamefont{M.~A.} \bibnamefont{Bedau}},
  \bibinfo{journal}{Science} \textbf{\bibinfo{volume}{303}},
  \bibinfo{pages}{963} (\bibinfo{year}{2004}).

\bibitem[{\citenamefont{Sinervo and Lively}(1996)}]{sinervo_n96}
\bibinfo{author}{\bibfnamefont{B.}~\bibnamefont{Sinervo}} \bibnamefont{and}
  \bibinfo{author}{\bibfnamefont{C.~M.} \bibnamefont{Lively}},
  \bibinfo{journal}{Nature} \textbf{\bibinfo{volume}{380}},
  \bibinfo{pages}{240} (\bibinfo{year}{1996}).

\bibitem[{\citenamefont{Burrows and Hawkins}(1998)}]{burrows_mep98}
\bibinfo{author}{\bibfnamefont{M.~T.} \bibnamefont{Burrows}} \bibnamefont{and}
  \bibinfo{author}{\bibfnamefont{S.~J.} \bibnamefont{Hawkins}},
  \bibinfo{journal}{Mar. Ecol. Prog} \textbf{\bibinfo{volume}{167}},
  \bibinfo{pages}{1} (\bibinfo{year}{1998}).

\bibitem[{\citenamefont{Szab{\'o} and
  Cz{\'a}r{\'a}n}(2001{\natexlab{a}})}]{szabo_pre01a}
\bibinfo{author}{\bibfnamefont{G.}~\bibnamefont{Szab{\'o}}} \bibnamefont{and}
  \bibinfo{author}{\bibfnamefont{T.}~\bibnamefont{Cz{\'a}r{\'a}n}},
  \bibinfo{journal}{Phys. Rev. E} \textbf{\bibinfo{volume}{63}},
  \bibinfo{pages}{061904} (\bibinfo{year}{2001}{\natexlab{a}}).

\bibitem[{\citenamefont{Nowak}(2006)}]{nowak06}
\bibinfo{author}{\bibfnamefont{M.~A.}~\bibnamefont{Nowak}},
  \emph{\bibinfo{title}{Evolutionary Dynamics: Exploring the Equations of Life}}
  (\bibinfo{publisher}{Harward University Press},
  \bibinfo{address}{Harvard}, \bibinfo{year}{2006}).

\bibitem[{\citenamefont{Hauert et~al.}(2002)\citenamefont{Hauert, De~Monte,
  Hofbauer, and Sigmund}}]{hauert_s02}
\bibinfo{author}{\bibfnamefont{C.}~\bibnamefont{Hauert}},
  \bibinfo{author}{\bibfnamefont{S.}~\bibnamefont{De~Monte}},
  \bibinfo{author}{\bibfnamefont{J.}~\bibnamefont{Hofbauer}}, \bibnamefont{and}
  \bibinfo{author}{\bibfnamefont{K.}~\bibnamefont{Sigmund}},
  \bibinfo{journal}{Science} \textbf{\bibinfo{volume}{296}},
  \bibinfo{pages}{1129} (\bibinfo{year}{2002}).

\bibitem[{\citenamefont{Traulsen and Claussen}(2004)}]{traulsen_pre04}
\bibinfo{author}{\bibfnamefont{A.}~\bibnamefont{Traulsen}} \bibnamefont{and}
  \bibinfo{author}{\bibfnamefont{J.~C.} \bibnamefont{Claussen}},
  \bibinfo{journal}{Phys. Rev. E} \textbf{\bibinfo{volume}{70}},
  \bibinfo{pages}{046128} (\bibinfo{year}{2004}).

\bibitem[{\citenamefont{Semmann et~al.}(2003)\citenamefont{Semmann, Krambeck,
  and Milinski}}]{semmann_n03}
\bibinfo{author}{\bibfnamefont{D.}~\bibnamefont{Semmann}},
  \bibinfo{author}{\bibfnamefont{H.-J.} \bibnamefont{Krambeck}},
  \bibnamefont{and} \bibinfo{author}{\bibfnamefont{M.}~\bibnamefont{Milinski}},
  \bibinfo{journal}{Nature} \textbf{\bibinfo{volume}{425}},
  \bibinfo{pages}{390} (\bibinfo{year}{2003}).

\bibitem[{\citenamefont{Tainaka}(1989)}]{tainaka_prl89}
\bibinfo{author}{\bibfnamefont{K.}~\bibnamefont{Tainaka}},
  \bibinfo{journal}{Phys. Rev. Lett.} \textbf{\bibinfo{volume}{63}},
  \bibinfo{pages}{2688} (\bibinfo{year}{1989}).

\bibitem[{\citenamefont{Frachebourg et~al.}(1996)\citenamefont{Frachebourg,
  Krapivsky, and Ben-Naim}}]{frachebourg_pre96}
\bibinfo{author}{\bibfnamefont{L.}~\bibnamefont{Frachebourg}},
  \bibinfo{author}{\bibfnamefont{P.~L.} \bibnamefont{Krapivsky}},
  \bibnamefont{and} \bibinfo{author}{\bibfnamefont{E.}~\bibnamefont{Ben-Naim}},
  \bibinfo{journal}{Phys. Rev. E} \textbf{\bibinfo{volume}{54}},
  \bibinfo{pages}{6186} (\bibinfo{year}{1996}).

\bibitem[{\citenamefont{Frachebourg and Krapivsky}(1998)}]{frachebourg_jpa98}
\bibinfo{author}{\bibfnamefont{L.}~\bibnamefont{Frachebourg}} \bibnamefont{and}
  \bibinfo{author}{\bibfnamefont{P.~L.} \bibnamefont{Krapivsky}},
  \bibinfo{journal}{J. Phys. A} \textbf{\bibinfo{volume}{31}},
  \bibinfo{pages}{L287} (\bibinfo{year}{1998}).

\bibitem[{\citenamefont{Szab{\'o} and Sznaider}(2004)}]{szabo_pre04a}
\bibinfo{author}{\bibfnamefont{G.}~\bibnamefont{Szab{\'o}}} \bibnamefont{and}
  \bibinfo{author}{\bibfnamefont{G.~A.} \bibnamefont{Sznaider}},
  \bibinfo{journal}{Phys. Rev. E} \textbf{\bibinfo{volume}{69}},
  \bibinfo{pages}{031911} (\bibinfo{year}{2004}).

\bibitem[{\citenamefont{He et~al.}(2005)\citenamefont{He, Cai, Wang, and
  Pan}}]{he_ijmpc05}
\bibinfo{author}{\bibfnamefont{M.}~\bibnamefont{He}},
  \bibinfo{author}{\bibfnamefont{Y.}~\bibnamefont{Cai}},
  \bibinfo{author}{\bibfnamefont{Z.}~\bibnamefont{Wang}}, \bibnamefont{and}
  \bibinfo{author}{\bibfnamefont{Q.-H.} \bibnamefont{Pan}},
  \bibinfo{journal}{Int. J. Mod. Phys. C} \textbf{\bibinfo{volume}{16}},
  \bibinfo{pages}{1861} (\bibinfo{year}{2005}).

\bibitem[{\citenamefont{Szab{\'o} and
  Cz{\'a}r{\'a}n}(2001{\natexlab{b}})}]{szabo_pre01b}
\bibinfo{author}{\bibfnamefont{G.}~\bibnamefont{Szab{\'o}}} \bibnamefont{and}
  \bibinfo{author}{\bibfnamefont{T.}~\bibnamefont{Cz{\'a}r{\'a}n}},
  \bibinfo{journal}{Phys. Rev. E} \textbf{\bibinfo{volume}{64}},
  \bibinfo{pages}{042902} (\bibinfo{year}{2001}{\natexlab{b}}).

\bibitem[{\citenamefont{Szab{\'o}}(2005)}]{szabo_jpa05}
\bibinfo{author}{\bibfnamefont{G.}~\bibnamefont{Szab{\'o}}},
  \bibinfo{journal}{J. Phys. A: Math. Gen.} \textbf{\bibinfo{volume}{38}},
  \bibinfo{pages}{6689} (\bibinfo{year}{2005}).

\bibitem[{\citenamefont{Frean and Abraham}(2001)}]{frean_prsb01}
\bibinfo{author}{\bibfnamefont{M.}~\bibnamefont{Frean}} \bibnamefont{and}
  \bibinfo{author}{\bibfnamefont{E.~D.} \bibnamefont{Abraham}},
  \bibinfo{journal}{Proc. R. Soc. Lond. B} \textbf{\bibinfo{volume}{268}},
  \bibinfo{pages}{1} (\bibinfo{year}{2001}).

\bibitem[{\citenamefont{Szab{\'o} and Szolnoki}(2002)}]{szabo_pre02a}
\bibinfo{author}{\bibfnamefont{G.}~\bibnamefont{Szab{\'o}}} \bibnamefont{and}
  \bibinfo{author}{\bibfnamefont{A.}~\bibnamefont{Szolnoki}},
  \bibinfo{journal}{Phys. Rev. E} \textbf{\bibinfo{volume}{65}},
  \bibinfo{pages}{036115} (\bibinfo{year}{2002}).

\bibitem[{\citenamefont{Sato et~al.}(2002)\citenamefont{Sato, Yoshida, and
  Konno}}]{sato_amc02}
\bibinfo{author}{\bibfnamefont{K.}~\bibnamefont{Sato}},
  \bibinfo{author}{\bibfnamefont{N.}~\bibnamefont{Yoshida}}, \bibnamefont{and}
  \bibinfo{author}{\bibfnamefont{N.}~\bibnamefont{Konno}},
  \bibinfo{journal}{Appl. Math. Comp.} \textbf{\bibinfo{volume}{126}},
  \bibinfo{pages}{255} (\bibinfo{year}{2002}).

\bibitem[{\citenamefont{Dickman}(1987)}]{dickman_pla87}
\bibinfo{author}{\bibfnamefont{R.}~\bibnamefont{Dickman}},
  \bibinfo{journal}{Phys. Lett. A} \textbf{\bibinfo{volume}{122}},
  \bibinfo{pages}{463} (\bibinfo{year}{1987}).

\bibitem[{\citenamefont{Szab{\'o} and F{\'a}th}(2006)}]{szabo_cm06}
\bibinfo{author}{\bibfnamefont{G.}~\bibnamefont{Szab{\'o}}} \bibnamefont{and}
  \bibinfo{author}{\bibfnamefont{G.}~\bibnamefont{F{\'a}th}}
  (\bibinfo{year}{2006}), \eprint{arXiv:cond-mat/0607344}.

\bibitem[{\citenamefont{Silva and Fontanari}(1999)}]{silva_epjb99}
\bibinfo{author}{\bibfnamefont{A.~T.~C.}~\bibnamefont{Silva}} \bibnamefont{and}
  \bibinfo{author}{\bibfnamefont{J.~F.}~\bibnamefont{Fontanari}},
  \bibinfo{journal}{Eur. Phys. J. B} \textbf{\bibinfo{volume}{7}},
  \bibinfo{pages}{385} (\bibinfo{year}{1999}).

\bibitem[{\citenamefont{Malcai et~al.}(2002)\citenamefont{Malcai, Biham,
  Richmond, and Solomon}}]{malcai_pre02}
\bibinfo{author}{\bibfnamefont{O.}~\bibnamefont{Malcai}},
  \bibinfo{author}{\bibfnamefont{O.}~\bibnamefont{Biham}},
  \bibinfo{author}{\bibfnamefont{P.}~\bibnamefont{Richmond}}, \bibnamefont{and}
  \bibinfo{author}{\bibfnamefont{S.}~\bibnamefont{Solomon}},
  \bibinfo{journal}{Phys. Rev. E} \textbf{\bibinfo{volume}{66}},
  \bibinfo{pages}{031102} (\bibinfo{year}{2002}).

\bibitem[{\citenamefont{Reichenbach et~al.}(2006)\citenamefont{Reichenbach,
  Mobilia, and Frey}}]{reichenbach_pre06}
\bibinfo{author}{\bibfnamefont{T.}~\bibnamefont{Reichenbach}},
  \bibinfo{author}{\bibfnamefont{M.}~\bibnamefont{Mobilia}}, \bibnamefont{and}
  \bibinfo{author}{\bibfnamefont{E.}~\bibnamefont{Frey}},
  \bibinfo{journal}{Phys. Rev. E} \textbf{\bibinfo{volume}{74}},
  \bibinfo{pages}{051907} (\bibinfo{year}{2006}).

\bibitem[{\citenamefont{Kowalik et~al.}(2002)\citenamefont{Kowalik, Lipowski,
  and Ferreira}}]{kowalik_pre02}
\bibinfo{author}{\bibfnamefont{M.}~\bibnamefont{Kowalik}},
  \bibinfo{author}{\bibfnamefont{A.}~\bibnamefont{Lipowski}}, \bibnamefont{and}
  \bibinfo{author}{\bibfnamefont{A.~L.} \bibnamefont{Ferreira}},
  \bibinfo{journal}{Phys. Rev. E} \textbf{\bibinfo{volume}{66}},
  \bibinfo{pages}{066107} (\bibinfo{year}{2002}).

\end{thebibliography}
\end{document}